# Formation of requirements traceability in the process of information systems design

Grigory Tsiperman [0000-0001-7740-5629]

Independent expert, Sol Liptsin 4/4, Jerusalem 9322403, Israel

`gntsip@gmail.com`

**Abstract.** The traceability of requirements in the information system design process is considered an important property of the project, one of its quality characteristics. The point here is that traceability provides the processes of validation and verification of software systems, and that the system model based on requirements traceability reduces the system's dependence on developers and, in general, makes it as clear as possible. One of the challenges of the traceability process, dubbed 'The grand challenge of traceability' among traceability researchers, is its integration into the design process. In this paper, to achieve this goal, we propose the application of the Adaptive Clustering Method (ACM) of Information System developed by the author, which is based on the idea of a seamless system architecture that provides explicit interconnection of project artifacts of different levels of abstraction.

## 1      Introduction

The information system design is an evolving, constantly refined system of requirements - from the users' business requirements to the requirements for the architecture and software components of the system. In other words, the information system project describes interrelated, consistent artifacts representing the system's requirements. It reflects different points of view on the information system, corresponding to different levels of abstraction of its perception by stakeholders (customer, analyst, architect, etc.).

The most effective criterion for assessing the quality of design before software code development is requirements tracing, which is defined as 'the ability to describe and trace the life cycle of a requirement in both forward and backward directions (i.e., from its origin, through its development and specification, to its subsequent deployment and use, and through periods of continuous refinement and iteration at any of these stages)' [1].

In terms of this approach to design, it seems surprising to learn from research ( [2], [3], [4]) that most practitioners do not even understand the term 'traceability.' Admittedly, this does not apply to practitioners working in regulated and/or safety-critical areas who have a much more complete understanding of what tracing does and how it should be installed. However, even in this case, as noted in [5], this does not mean that they implement traceability more readily or effectively: analysis of traceability documents revealed numerous problems with the completeness and correctness of traceability data [6].

Tracing of requirements is generally not integrated into the design process. Paper [5] articulated a major problem associated with traceability: the cost, effort, and discipline required to create and maintain traceability in a rapidly evolving software system can be extremely high.

This paper attempts to propose a design-integrated tracing generation process based on the Adaptive Clustering Method (ACM) of information systems ( [7], [8]). ACM is an MDD method where detailed architectural models are built based on the decomposition of elements of models of a higher level of abstraction, ensuring their seamless connectivity, i.e., such connectivity that the ability to trace transitions between elements of architectural models is not compromised during decomposition.

This paper has the following structure: Section 2 provides a brief literature review on the problems of requirements tracing. Section 3 formulates the problems whose solution should provide traceability of design artifacts. Sections 4 and 5 discuss the solutions to these problems. Section 7 concludes the paper.

## 2      Related works

In this paper, system requirements tracing is divided into two main groups with respect to Requirement Specification (RS): tracing stakeholder requirements obtained during the research phase of the system's subject area (pre-RS), and requirements generated during system design from RS to code requirements (post-RS).

### 2.1      Pre-RS Traceability

Pre-RS tracing refers to the definition of the part of the requirements system that is formed before the system requirements specification (RS) is created; that is, the task of pre-RS tracing is to make the system requirements specification.

The paper [9] provides a detailed review of research results regarding pre-RS requirements.

The source of the system requirements is artefacts (documents, text fragments, media data) obtained during the research, linked by trace references to the RS. Tracing is necessary (a) to justify the system requirement in terms of identifying the source (stakeholder, rule artefact or constraints); (b) to analyze the consistency of requirements in the specification (cross-tracing of requirements [10]; e.g. if performance requirements contradict system cost constraints).

However, the cost required for pre-RS tracing is often considered too high (see [1], [11]). In my practice, I have encountered projects where not only the sources of the requirement's origin were not traced, but even their specification (RS) was not formed as an artefact: each requirement was discussed during a scram sprint and went straight to the analyst for implementation of a separate statement and then to the programmer for coding. As a result, the trace included three artefacts: the requirement for implementation and the codes of the backend and frontend parts of the system. Of course, when the question of further development of IS development arose, the requirements specification was collected, but there was no trace of the origin of the requirements.

The main consequences of inadequate pre-RS traceability, as stated in [9], can result in the system not meeting the stakeholders' expectations. Such consequences relate to the following aspects:

- "*Black box*" requirements - without any communication details with post-RS. The lack of such details becomes more critical in long-term projects, especially if the responsible employees change.
- *Expensive consequences beyond the time frame* - if requirements problems are not identified during creation and maintenance before implementation, it becomes more expensive to resolve them later.

Keeping requirements up to date and understanding their context is essential to the success of an evolving project when it becomes increasingly complex.

## 2.2 Post-RS Traceability

Traceability after specification (post-RS) is the ability to link RS requirements to project artifacts. The *Centre of Excellence for Software Traceability* (CoECT) 's research on post-RS traceability is focused on this area [12].

Since the path from the requirement in RS to the software code may not be close, passing through several levels of abstraction, obtaining traceability becomes non-trivial. Over the past decades, techniques have been developed that aim to create trace links by analyzing the textual content of each software artifact, including code. Many variants and modifications have been proposed to improve traditional term-based trace reference retrieval approaches (see [13], [14]). Deep learning techniques have been applied to improve the accuracy of trace references [15]. However, as noted in [16], applying deep learning leads to opaque results for safety-critical systems, i.e., one cannot inspect the code or conduct exhaustive testing.

There is some distrust of the concept of tracing. Approaches that have proven their value in theory are perceived as too theoretical and not applicable [17]. However, the practical usefulness of tracing has been verified in several experiments. Paper [18] shows that teams using tracing were, on average, 24% faster at completing a given task and made, on average, 50% more correct decisions. One of the main results of this study was the proof that tracing significantly affects defect rates.

Paper [5] identifies target perspectives for solving the traceability problem, which should lead to a final solution called '*The grand challenge of traceability*' [18]. This initiative is formulated as follows: "Traceability is always there; you don't have to think about getting it because it is built into the engineering process; traceability has effectively 'disappeared without a trace.'"

An intuitive understanding of the usefulness of tracing is likely to be present among specialists, but focusing on the main tasks of projects forces them to ignore tracing, especially since the significant results of tracing will clearly appear only at the stage of system development and operation. It is true that in critical areas the presence of traceability allows for painless certification, however, as noted in [5]: ...Traceability data was not only incomplete, incorrect, and confusing in many cases, but there were clear indications that traceability links were created late in the process in many projects, especially for certification processes.

## 3     Problem Statement

Broadly, the definition of traceability extends to the scope and relationship of any uniquely identifiable systems engineering artefact to a wide range of system-level components such as people, processes and hardware models (*Software and Systems Traceability*, [19]). All these components represent objects subject to requirements in the design of an information system. In other words, requirements traceability concerns tracing design artefacts relating to information system objects. This fact likely allowed us to formulate the concept of 'The grand challenge of traceability': traceability is always present, and one does not even have to think about achieving it because it is embedded in the design process...' [19].

As noted earlier, information system design can be seen as a process of specifying requirements at different levels of abstraction since the project artefacts being formed relate to the future system and must be implemented in code.

The practice of information systems architecture design involves a heuristic transition between architectural models of different levels of abstraction: a more detailed architecture is built so that technical solutions correspond to the model of the previous level as much as possible. In this case, the architect uses his creativity, experience and knowledge to build a detailed model and then proves (or considers it obvious) that the resulting model satisfies the requirements derived from the more abstract architecture. This aspect of information systems design has been emphasized in [20]. Let us denote it as a *technological gap* between architectural models of different levels of abstraction. The technological gap leads to the fact that the architectural artefacts of the project belonging to different levels of abstract are not explicitly linked, i.e. tracing between them is impossible.

**The first task** to be solved is to propose a way to bridge the gap and build a *seamless system architecture* in which tracing is possible for project artefacts.

**The second task** is to define *types of architectural elements belonging to architectural models of different levels of abstraction* that provide tracing of design artefacts from stakeholder requirements (pre-RS) to software code (post-RS).

These problems are solved by a seamless architecture design technique based on the application of the Adaptive Clustering Method (ACM, see [7], [8]), in which detailed architectural models are built based on inference from high-level abstraction models, ensuring that they are seamlessly connected and traceable between their constituent elements.

## 4 Seamless architecture of an information system

An architectural description of a system is a collection of architectural views corresponding to different viewpoints on the system. Consider **Fig. 1**, which is a fragment of the conceptual model of an architectural view from [21], supplemented with an element defining a seamless connection between architectural models.

An architectural view corresponds to a certain viewpoint on the architecture and defines the system's architecture with a level of detail corresponding to the stakeholder's interests. In the Zachman scheme, for example, the viewpoints correspond to the participants in the system creation process: planner, analyst, architect, designer, programmer, and operator. Considered in this sequence, these viewpoints lead to related architectural views of an increasingly higher level of detail.

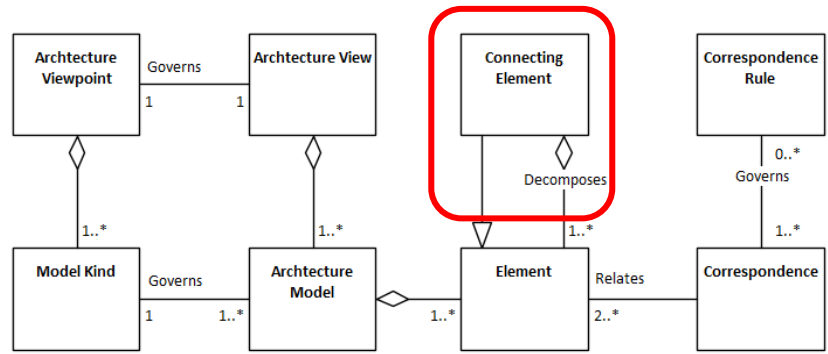

**Fig. 1.** Conceptual model of a seamless architecture.

Among the elements of architectural models that define abstracts included in the architectural representation, elements (*Connecting Elements*) are defined, the decomposition of which defines elements of architectural models of the next level of abstraction. For example, such an element can be a business function decomposed into more detailed functional elements, or a software module of a subsystem decomposed into components corresponding to the program code. Thus, Connecting Element explicitly defines the connections between elements of architectural models, including those related to different levels of abstraction. This is the idea of seamless architecture: design artifacts are connected based on the decomposition of elements of architectural models.

## 5 Types of architectural elements providing tracing

As follows from the previous section, connecting elements included in the architectural models of architectural views of the information system can be used as elements that provide tracing. To determine which elements should be used, let us consider the architectural views of the information system related to different levels of abstraction: Business architecture, Functional architecture, Component architecture, Data architecture. Fig. 2 presents a model of the seamless connection of these architectures. This section explains the figure.

Since the description is sufficiently abstract, it is suggested that you familiarize yourself with the paper's example of using ACM to design an information system for a more detailed understanding [8].

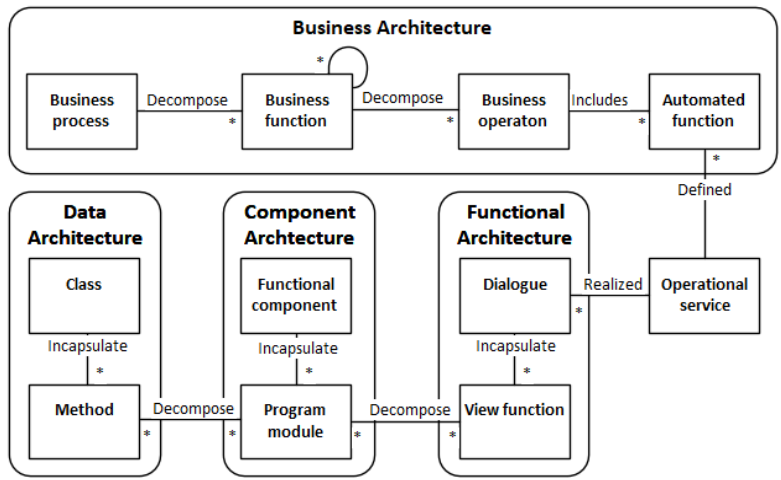

**Fig. 2.** Seamless interconnection of architectural representations of the information system

### 5.1   Business architecture of information systems

In ACM, business architecture is the starting point for information system design. An information system's business architecture is a set of enterprise business process models represented by UML *Activity Diagrams* (Fig. 3).

In the context of this work, the target business architecture is considered, i.e., business process models that consider the use of the information system.

Models of business architecture are built on three main elements (Fig. 2) defined in the decomposition procedure: (a) *Business process*, understood in the usual sense, is a decomposable action that transforms input objects [resources] into a result valuable for the client; (b) *Business function* is an action from a set of actions determined by the decomposition of a business process or another business function; (c) *Business operation* is a business function that is not decomposed within the framework of the description of a business process, the executor of which is a specific employee.

   Decomposition defines explicit links between these architectural elements and is completed when all business functions are decomposed into business operations. Thus, the connecting element in business architecture is an action presented as a business function.

A business operation is where a user interacts with an information system. It includes user functions, some of which need to be automated (*automated functions*).

Business architecture is built based on the results of an analytical study of the enterprise's business processes. Specific survey material must substantiate each fact reflected in the business architecture model. Such substantiation is a tracing link between the original survey artifact (e.g., a text or media file with the results of analytical interviews, survey data on materials regulating the implementation of the business processes being studied, etc.) and the architectural element of the model (pre-RS tracing).

The description of each business architecture element must reflect business requirements and stakeholder requirements, which serve as the initial data for determining the functional and non-functional requirements for the information system and its individual functions.

The requirements defined in the business architecture model form the basis of the RS. Artifact tracing is a chain of traces, which are triplets of elements, including a source artifact, a target artifact, and a trace link [19]. Pre-RS tracing is a chain of traces from a business process through business functions to a business operation, where trace links are based on typed decomposition relationships between source and target artifacts.

### 5.2    Information system business operations services

Business architecture decomposes the future IS into clusters at the level of requirements for the automation of individual workplaces determined by business operations. The business architecture model ensures the consistency of workplace functioning and their interrelationships within the automated business process.

The transition from business architecture to system architecture (see Fig. 2) is defined through *business operation services* (*Operational Service*). A business operation service (service) is an abstract architectural object representing the requirements for automated functions that support executing the corresponding business operation. The service is connecting element between business architecture and the functional architecture of the information system.

Operation services are formed for each business operation, which includes automated functions and defines the boundary between the business architecture and the IS system architecture (Fig. 3).

An operation service defines the requirements specification for implementing the corresponding business operation and is a connecting architectural element that does not imply software implementation. It is defined based on traces, where the corresponding business operation is the final target element for the trace chain. The set of all operation services forms the RS.

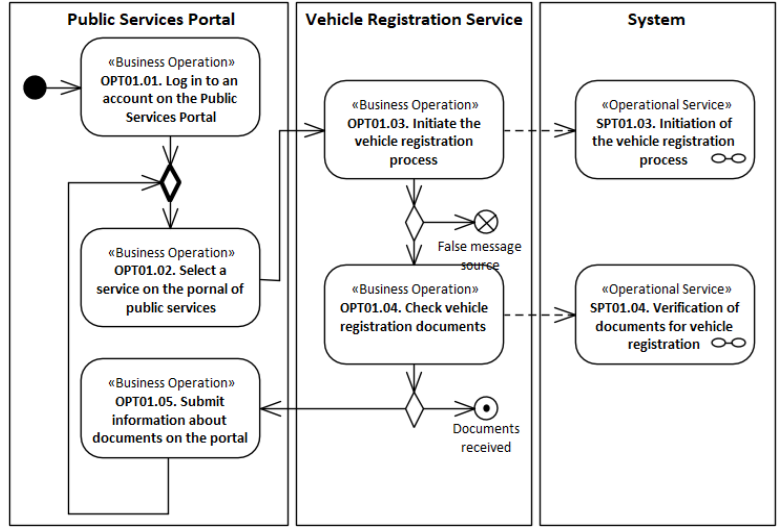

**Fig. 3.** Business process model with operational services

### 5.3   Functional architecture of information system

Functional architecture is an architectural representation that includes architectural models of the IS functions that implement the automated functions of business operations. In other words, functional architecture models the IS interaction with users and other external agents. The functional architecture forms the appearance of the IS based on the presentation of compositions of the system's dialogues, defines the requirements for dialogues, and specifies interfaces with external systems.

The main element of functional architecture is (Fig. 2): (a) *Dialogue* is an interaction between system agents that causes a change in the state of the IS by launching the corresponding software components [22]; (b) *View functions* are the functions of processing input/output data, changing the state of the IS, and servicing errors and restrictions implemented by the dialogue.

Dialogue is understood broadly as the user's interaction with the computer and the exchange of messages between any objects (agents) of the IS and external systems. The functional architecture model describes the *service scenario* as a logical combination of IS dialogs reflecting the order of their call to implement the automated functions of the business operation. The UML sequence diagram is also used to build a service scenario model in the ACM.

The functional architecture model describes a service scenario — a logical combination of information system dialogs, reflecting the order of their invocation to implement the business operation functions being automated. The UML sequence diagram is also used to build a service scenario model in ACM. The requirements included in a service form traces that include the initial service requirement, a scenario dialog and a trace reference.

The view functions define the requirements for the software implementation of the architectural elements of the next level—the modules of the component architecture, which represent the structural decomposition of the IS into functional components.

### 5.4 Component architecture of information system

Component architecture establishes the composition and interaction of the IS functional components, defines software modules and their distribution among components, and details the IS functional requirements.

The main elements of the component architecture model are (Fig. 2): (a) *Functional components* (subsystems and external IS) are determined by the choice of an architectural pattern (or set of architectural patterns) and the system's external environment; (b) *Software modules* are structural parts of the functional components of the IS implemented in the program code, detailing the view functions.

The connecting element between the functional and component architectures is the view function, decomposed into modules of the component architecture. This decomposition is performed on the functional components that encapsulate the software modules (Fig. 4). Therefore, Sequence Diagrams are used for the model of component architecture in the ACM, defining the functional components' modular composition.

The tracing is defined by traces containing the view function as the source artifact, the software module as the target artifact, and the corresponding trace link.

### 5.5 Data architecture of information system

The data architecture is seen as a set of attributed classes encapsulating the relevant methods, the requirements for which need to be defined (**Fig. 2**).

Like the component architecture, class methods are determined by decomposing software modules performed on a set of classes. For this, sequence diagrams are used, in which classes are lifelines. The decomposition of a software module can include previously defined class methods; there is a "many-to-many" relationship between modules and class methods.

For large systems with a complex data model, the question of which classes to decompose the software module on is non-trivial. As shown in [23], business operations divide the subject area of an information system into relatively weakly coupled clusters, which can form the basis of the system's microservices. In the context of this work, this means that the data architecture can be represented by weakly coupled data submodels corresponding to business operations or, equivalently, to operation services. This significantly limits the number of data model classes needed to decompose software modules to determine class methods - it is sufficient to build a backtrace of the following artifacts:

*software module → view function → dialog → operation service.*

The operation service uniquely defines the data submodel that should be used to decompose the software module.

Finally, the traceback for a class method includes a trace link with the software module as the source artifact and the class method as the target artifact.

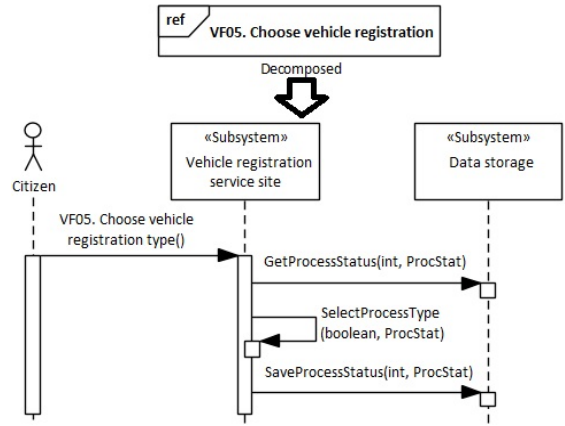

**Fig. 4.** Definition of IS component modules

## 6      Conclusion

ACM is a generalization of the author's more than 30 years of real-world experience designing various information systems and teaching the discipline of "System Analysis and IS Design" at the university.

Enterprise Architect by Sparx is currently used as the main design tool for ACM. The task is to adapt the tool to construct a traceability matrix and form individual trass. The formation of traces in the ACM is carried out within the framework of the IS design and, therefore, does not require a separate process of maintaining requirements. Also, the use of tools for the formation of requirement traces with this approach loses its meaning.

The system-forming element for requirements is tracing, which links requirements of different levels of abstraction, defining the place of each requirement in the system of requirements. It is not possible to compare ACM with other methods because no other similar approaches to integrating traceability with the IS design process are known, as evidenced by the CoECT (see for example [24]).

*The main contribution of this work* is the theoretical justification and practical application of the ACM for integrating the tracing of system artifacts into the process of designing an information system. This corresponds to the target perspective of tracing software and systems engineering development, designated in [19] as the "Grand Challenge of Traceability." This approach ensures the completeness of the functional implementation of the IS since the decomposition of the business process allows for the precise formulation of user requirements for automated functions. Further design of functional components at various levels of the abstract description of the information system is essentially a substantiated conclusion about the necessary and sufficient functionality of the IS, which allows for avoiding errors associated with technological gaps between architectural models and insufficient or excessive functionality.

The development directions of the ASM include the development of a reporting tool for requirements tracing, which allows for the automatic construction of a tracing matrix and individual arbitrary traces.

**Acknowledgments.** The author expresses sincere gratitude to Professor Boris Pozin, who provided invaluable assistance in preparing this article.

**Disclosure of Interests.** The authors have no competing interests to declare that are relevant to the content of this article.

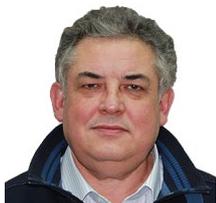
Expert in the system analysis and design of information systems with over 30 years of experience. Worked in leading Russian IT companies. Now, I am an independent expert. As a leading architect and technical manager, I have implemented many corporate systems projects and public administration projects, including the Russian system of biometric passports for citizens. The area of research interest is the design of information systems, microservice architecture, and project management - the author of more than 30 scientific publications. I teach the original "System Analysis and IS Design" course at the university, which is based on the Adaptive Clustering Method.